\documentclass[twocolumn,showpacs,preprintnumbers,amsmath,amssymb]{revtex4}

\usepackage{amsmath,amssymb}
\usepackage[latin1]{inputenc}
\usepackage{dcolumn}
\usepackage{bm}

\newcommand{\diff}[0]{\text{d}}
\newcommand{\im}[0]{\text{Im}\,}
\newcommand{\re}[0]{\text{Re}\,}
\usepackage{graphicx}
\usepackage{dcolumn}
\usepackage{bm}

\begin{document}


\title{Fresnel equations and the refractive index of active media}

\author{Johannes Skaar}
\affiliation{Department of Electronics and Telecommunications\\
Norwegian University of Science and Technology}

\date{\today}

\begin{abstract}
There exists a class of realizable, active media for which the refractive index cannot be defined as an analytic function in the upper half-plane of complex frequency. The conventional definition of the refractive index based on analyticity is modified such that it is valid for active media in general, and associated Fresnel equations are proved. In certain active media, the presence of a ``backward'' wave, for which both phase velocity and Poynting's vector point towards the excitation source, is demonstrated. 
\end{abstract}

\maketitle

\section{Introduction}
With the recent advances in the fabrication of passive materials and metamaterials \cite{smith, pendry2004}, it seems plausible that novel classes of active materials may be developed. While the basic electromagnetic concepts, such as microscopic and relativistic causality, certainly apply to active media, the analysis of electromagnetic wave propagation may differ. For example, at normal incidence to a boundary between two passive media, the transmitted field is proportional to $\exp(in\omega z/c)$, where $n$ is the refractive index, $\omega$ is the angular frequency, $z$ is the coordinate in the direction away from the boundary, and $c$ is the vacuum velocity of light. Since $z$ is arbitrary, relativistic causality dictates $n$ to be an analytic function in the upper half of the complex $\omega$-plane \cite{nussenzveig,brillouin}. On the other hand, at least in principle there exist realizable, active media for which $n$ cannot be identified as an analytic function in the upper half-plane. Indeed, consider a Lorentz model with relative permittivity $\epsilon=1-f$, where $f$ is given by
\begin{equation}
\label{lorentz}
f(\omega)=\frac{F\omega_{0}^2}{\omega_{0}^2-\omega^2-i\omega\Gamma}.
\end{equation}
Here, $\omega_{0}$, and $\Gamma$ are positive parameters. If $F>0$ this medium is active; furthermore, if $F>1$, $\epsilon$ has a simple zero in the upper half-plane. Thus, assuming the relative permeability $\mu=1$, when $F>1$ the refractive index $n=\sqrt{\epsilon}\sqrt{\mu}$ contains a branch point in the upper half-plane.

These media would seem to violate relativistic causality. This is certainly not the case; to resolve this apparent paradox, I will provide a rigorous derivation of the Fresnel equations in the general case, based on the electromagnetic solutions of a finite slab. The definition of the refractive index is generalized so that it is valid for active media. It will become clear that when the refractive index contains branch points in the upper half-plane, the Fresnel equations and the refractive index do not have unique physical meaning for real frequencies. The Kramers-Kronig relations for the refractive index must be modified accordingly.

As an extra bonus, the general framework in this paper can be used to analyze novel classes of active media. Recently, it has been suggested that certain active, nonmagnetic media with simultaneously passive and active material resonances can yield negative refraction \cite{chen2005}. I will consider the Fresnel equations for these types of materials, and demonstrate that at normal incidence, the ``backward'' wave is excited so that both phase velocity and Poynting's vector point towards the excitation source. While negative refraction is found at oblique incidence, it is shown that evanescent field amplification, similarly to that in the Veselago-Pendry lens \cite{veselago,pendry2000}, does not happen.

The remaining of the paper is structured as follows: In Section II the theory of the electromagnetic parameters is reviewed. Section III contains an electromagnetic analysis of a finite slab surrounded by vacuum. From the resulting fields, in Section IV the Fresnel equations are proved in the general case, and the associated refractive index is identified. The results are interpreted in the form of examples (Section V); inverted Lorentzian media, and more complex media featuring negative refraction.

\section{The electromagnetic parameters}
Any electromagnetic medium must be causal in the microscopic sense; the polarization and magnetization cannot precede the electric and magnetic fields, respectively. This means that the relative permittivity $\epsilon(\omega)$ and the relative permeability $\mu(\omega)$ obey the Kramers-Kronig relations. In terms of the susceptibilities $\chi=\epsilon-1$ or $\chi=\mu-1$, these relation can be written
\begin{subequations}
\label{KKchi}
\begin{align}
&\im \chi=\mathcal H\,\re \chi,\label{KKchi1}\\
&\re \chi=-\mathcal H\, \im \chi,\label{KKchi2}
\end{align}
\end{subequations}
where $\mathcal H$ denotes the Hilbert transform \cite{kronig,kramers,nussenzveig,landau_lifshitz_edcm}. These conditions are equivalent to the fact that $\chi$ is analytic in the upper half-plane ($\im\omega>0$), and uniformly square integrable in the closed upper half-plane (Titchmarsh' theorem) \cite{nussenzveig,titchmarsh}\footnote{If the medium is conducting at zero frequency, the electric $\chi$ is singular at $\omega=0$. Although $\chi$ is not square integrable in this case, similar relations as \eqref{KKchi} can be derived \cite{landau_lifshitz_edcm}.}. The susceptibilities are defined for negative frequencies by the symmetry relation
\begin{equation}
\chi(-\omega)=\chi^*(\omega), \label{sym}
\end{equation}
so that their inverse Fourier transforms are real. 

For passive media, in addition to \eqref{KKchi} and \eqref{sym} we have: 
\begin{equation}
\im \chi(\omega)>0 \text{ for } \omega>0. \label{loss}
\end{equation}
The losses, as given by the imaginary parts of the susceptibilities, can be vanishingly small; however they are always present unless we are considering vacuum \cite{landau_lifshitz_edcm}. Eqs. \eqref{KKchi}-\eqref{loss} imply that $\im\chi>0$ in the first quadrant ($\re\omega>0$ and $\im\omega\geq 0$), and that $1+\chi$ is zero-free in the upper half-plane \cite{landau_lifshitz_edcm}. Thus for passive media, the refractive index $n=\sqrt{\epsilon}\sqrt{\mu}$ can always be chosen as an analytic function in the upper half-plane. With the additional choice that $n\to +1$ as $\omega\to\infty$, $n$ is determined uniquely by 
\begin{equation}
\label{nfromepsilonmu}
n=\sqrt{|\epsilon||\mu|}\exp[i(\arg\epsilon+\arg\mu)/2],
\end{equation}
where the complex arguments are restricted to the interval $(-\pi,\pi]$. From Eq. \eqref{nfromepsilonmu} it is immediately found that \eqref{sym} and \eqref{loss} hold for the substitution $\chi\to n$. Moreover, the Kramers-Kronig relations for the refractive index are established using Titchmarsh' theorem by noting that $n-1$ is analytic in the upper half-plane, and that $n-1$ has the required asymptotic behavior: The fact that $\chi$ satisfies \eqref{KKchi} means that $\chi$ satisfies the square integrability condition. From $n=\sqrt{\epsilon}\sqrt{\mu}$, it follows that $n-1$ has a similar asymptotic behavior. This gives
\begin{subequations}
\label{KKn}
\begin{align}
&\im n   = \mathcal H\,\{\re n-1\},\label{KKn1}\\
&\re n-1 =-\mathcal H\, \im n.\label{KKn2}
\end{align}
\end{subequations}

On the basis of causality, it is clear that the susceptibilities of active media also satisfy \eqref{KKchi} and \eqref{sym}. However, since \eqref{loss} does not hold for active media, it is no longer true that $\epsilon$ and $\mu$ are zero-free in the upper half-plane. Indeed, we have already seen that the permittivity associated with an inverted Lorentz model may have a simple zero in the upper half-plane. Thus, the refractive index cannot always be chosen as an analytic function in the upper half-plane.

To clarify this point, I will consider one-dimensional electromagnetic wave propagation in an active medium. The main goal is to identify the reflection and transmission at a single interface, and to prove that these waves obey relativistic causality. The Fresnel equations could be proved in the conventional way; however, it might not be clear {\it a priori} how to specify the wavevector of the transmitted wave in the active medium. In fact, contrary to the case with passive media, it is not necessarily correct to let the Poynting vector of the transmitted wave point away from the interface. Moreover, it is not always possible to choose the wavevector as an analytic function for $\im\omega>0$. Therefore, the active medium is assumed to have finite thickness initially. Then the transmitted wavevector at the far end of the slab can be specified trivially, and the electromagnetic solutions can be found. Subsequently, using the result of the finite slab, the fields of a half-infinite medium can be determined.

\section{Laplace transform analysis of an active slab}
\begin{figure}
\includegraphics[height=4.5cm,width=4.5cm]{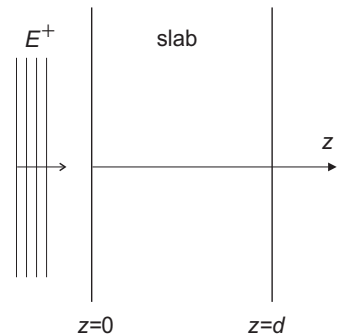}
\caption{A plane wave incident to a slab of thickness $d$.}
\label{fig:slab}
\end{figure}

Consider a plane wave in vacuum, normally incident to a linear, isotropic, and homogenous slab with parameters $\epsilon$ and $\mu$, see Fig. \ref{fig:slab}. While vacuum is chosen as the surrounding medium, it should be clear that any other passive media yield similar results. The slab is located in the interval $0\leq z\leq d$, where $d$ is the thickness of the slab. A causal excitation on the left-hand side of the slab ($z=0^-$) is assumed, i.e., the fields for $z\geq 0$ vanish for time $t\leq 0$. Since the medium is active, the physical, time-domain fields may diverge as $t\to\infty$, so Fourier transformed fields do not necessarily exist. (Strictly speaking, in practical materials, the fields do not diverge. When the fields become sufficiently strong, they deplete the gain. However, we restrict ourselves to the linear regime of the susceptibilities, and bear in mind that the model breaks down when the fields become large. The divergences are indicators of instabilities such as, for example, the ignition of lasing.) Thus, instead of Fourier transforms, we use Laplace transforms of the fields. Denoting the physical, time-domain electric field $\mathcal E(z,t)$, the Laplace transformed field for $z\geq 0$ is
\begin{equation}
\label{laplzgeq0}
E(z,\omega)=\int_0^\infty \mathcal E(z,t)\exp(i\omega t)\diff t,
\end{equation}
where $\im\omega\geq\gamma$, for a sufficiently large, real constant $\gamma$. Assuming that the source is located at $z=z_{\text{s}}<0$, \eqref{laplzgeq0} still applies for $z_{\text{s}}<z<0$ provided the lower limit in the integral is replaced by $z_{\text{s}}/c$. Since the fields everywhere are assumed to vanish for $t$ equal to the lower limit in the integral, the partial time derivatives in Maxwell's equations become simply $-i\omega$ in the transform domain. At the boundaries the Laplace transformed electric and magnetic fields must be continuous. Thus we can use a similar analysis method as in the Fourier case, considering the transformed fields at each $\omega$. Assuming the forward propagating electric field $\mathcal E^+(z,t)$ for $z<0$, and recalling that $\mathcal E(z,t)$ for $z>d$ is assumed to vanish for negative time, we can write
\begin{eqnarray}
&& E(z,\omega)/E^+(0,\omega) \\
&& = 
\begin{cases}
\exp(i\omega z/c)+R\exp(-i\omega z/c) & \text{for $z_{\text{s}}<z<0$},\\
S^+\exp(ikz)+S^-\exp(-ikz) & \text{for $0\leq z\leq d$},\\
T\exp(i\omega z/c) & \text{for $z > d$},
\end{cases} \nonumber
\end{eqnarray}
where $E^+(0,\omega)$ is the Laplace transform of the excitation $\mathcal E^+(0,t)$. By matching the electric and magnetic fields at both interfaces, the total reflection coefficient $R$, field amplitudes in the slab ($S^+$ and $S^-$), and transmission coefficient $T$ are found to be
\begin{subequations}
\label{slabrt}
\begin{align}
&R=\frac{(\eta^2-1)\exp(-ikd)-(\eta^2-1)\exp(ikd)}{(\eta+1)^2\exp(-ikd)-(\eta-1)^2\exp(ikd)}, \label{slabr}\\
&S^+=\frac{2\eta(\eta+1)}{(\eta+1)^2-(\eta-1)^2\exp(2ikd)}, \label{slabsp}\\
&S^-=\frac{2\eta(\eta-1)}{(\eta-1)^2-(\eta+1)^2\exp(-2ikd)},\label{slabsm}\\
&T=\frac{4\eta}{(\eta+1)^2\exp(-ikd)-(\eta-1)^2\exp(ikd)}.\label{slabt}
\end{align}
\end{subequations}
Here, $k=n\omega/c$, $\eta=\mu/n$, and $n=\sqrt{\epsilon\mu}$. Note that $R$, $S\equiv S^+\exp(ikz)+S^-\exp(-ikz)$, and $T$ are unchanged if $n\to-n$, so the choice of the sign of $n$ does not matter. In other words, only even powers of $n$ are present in $R$, $S$, and $T$, which means that the fields contain no branch points in the upper half-plane.

For active media, $R$, $S$, and $T$ may contain poles in the upper half-plane.
Despite poles, the fields are relativistically causal. This fact is established using the analyticity of the respective functions in the half-plane $\im\omega>\gamma$ and their asymptotic behavior in that region. For example, $T$ tends to $\exp(i\omega d/c)$ as $\im\omega\to\infty$, which means that the signal-front delay through the slab is $d/c$ \footnote{\label{signalfrontdelay}Let $F(s)$ be the Laplace transform of the Laplace transformable function $f(t)$. A well-known result in the theory of Laplace transforms states that if $F(-i\omega)\exp(-i\omega\tau)\to 0$ as $\im\omega\to\infty$, then the inverse transform of $F$ is zero for $t<\tau$ (See for example \cite{papoulis}).}.

The time-domain solutions are found by inverse Laplace transforms of the associated fields. For example, if the excitation is $\mathcal E^+(0,t)=u(t)\exp(-i\omega_1 t)$, where $u(t)$ is the unit step function and $\im\omega_1<0$, the time-domain field in the slab is \footnote{Any physical excitation is real and can be represented by two complex terms, e.g., $u(t)\exp(-i\omega_1 t)+u(t)\exp(i\omega_1^* t)$. This excitation would require the substitution $1/(i\omega_1-i\omega)\to 1/(i\omega_1-i\omega)+1/(-i\omega_1^*-i\omega)$ in \eqref{invlapl}. For clarity it is often convenient to consider the positive frequency excitation separately.}
\begin{equation}
\label{invlapl}
\mathcal E(z,t)=\frac{1}{2\pi}\int_{i\gamma-\infty}^{i\gamma+\infty} \frac{S}{i\omega_1-i\omega}\exp(-i\omega t)\diff\omega.
\end{equation}
When $S$ has no poles in the closed upper half-plane, $\gamma$ can be set to zero. It follows then that the field $S$ can be interpreted in the usual way at real frequencies. When $S$ contains poles in the upper half-plane, the integration path in \eqref{invlapl} should be located above all of them. If we still insist on locating the integration path along the real axis, the resulting discrepancy would correspond to the sum of residues of the poles in the upper half-plane. Such residues are exponentially increasing and demonstrate clearly the fact that when $S$ contains poles in the upper half-plane, the slab is electromagnetically unstable.

\section{Fresnel equations}
To facilitate the analysis and interpretation of active media, it is useful to obtain the fields when the slab fills the entire half-plane $z\geq 0$. For passive media, one can take the limit $d\to\infty$ in \eqref{slabrt} to obtain
\begin{subequations}
\label{fresnel}
\begin{align}
&R=\frac{\eta-1}{\eta+1},\label{fresnelr}\\
&S=\frac{2\eta}{\eta+1}\exp(in\omega z/c). \label{fresnels}
\end{align}
\end{subequations}
In obtaining \eqref{fresnel} the sign of $n$ has been chosen such that $\im n$ is positive in the first quadrant ($\re\omega>0$ and $\im\omega\geq 0$). For passive media, this choice is equivalent to the choice in \eqref{nfromepsilonmu}, or in the conventional definition of $n$ based on analyticity. Eqs. \eqref{fresnel} are essentially the Fresnel equations; however the propagation factor $\exp(in\omega z/c)$ has been inluded in the transmitted field, to emphasize the position dependence of the field in the active medium.

For active media, one could guess that \eqref{fresnel} remains valid, but it is not clear {\it a priori} how to choose $n$. The imaginary parts of $\epsilon$ and $\mu$ may take both signs in the first quadrant, and it may not be possible to choose $n$ as an analytic function in the upper half-plane. Surprisingly, as shown by the following example, taking the limit $d\to\infty$ in \eqref{slabrt} leads in general to unphysical results. Consider a slab of an inverted Lorentzian medium $\epsilon=1-f$ and $\mu=1$, where $f$ is given by \eqref{lorentz}. Assuming $F$ is small (corresponding to low gain), $n\approx \pm(1-f/2)$. In the limit $d\to\infty$, we find $R=(\eta+1)/(\eta-1)$ and $S=2\eta\exp(-in\omega z/c)/(\eta-1)$, where $n=1-f/2$. This solution is clearly unphysical as $R,S\to\infty$ as $\omega\to\infty$; a convergence half-plane $\im\omega\geq\gamma$ does not exist. 

To find the correct solution, we start with the time-domain solution \eqref{invlapl}, where $d$ is finite. Due to the finite bandwidth of physical media, $|\chi|\propto 1/\omega^2$ as $\omega\to\infty$ in the upper half-plane \cite{nussenzveig}. Thus, for a sufficiently large $|\omega|$, $\epsilon\approx\mu\approx 1$. Choosing the sign of $n$ such that $n\approx 1$ here leads to $|n-1|\propto 1/\omega^2$ for $\omega\to\infty$. This implies in turn that there exists a $\gamma>0$ such that $|[(\eta-1)/(\eta+1)]\exp(ikd)|<1$ for any $\omega$ with $\im\omega\geq\gamma$. We can now expand \eqref{slabsp} and \eqref{slabsm} into geometrical series in $[(\eta-1)/(\eta+1)]^2\exp(2ikd)$. Hence, $S\equiv S^+\exp(ikz)+S^-\exp(-ikz)$ becomes
\begin{eqnarray}
S&&=\frac{2\eta e^{ikz}}{\eta+1}\left[1+\left(\frac{\eta-1}{\eta+1}\right)^2 e^{2ikd}+\ldots \right] \\
&&-\frac{2\eta(\eta-1)e^{2ikd-ikz}}{(\eta+1)^2}\left[1+\left(\frac{\eta-1}{\eta+1}\right)^2 e^{2ikd}+\ldots \right] \nonumber
\end{eqnarray}
After substitution of $S$ into \eqref{invlapl}, and using the signal-front delay property of the Laplace transform [14], it is apparent that only the first term leads to a nonzero result for $t<2d-z$. This term coincide with \eqref{fresnels}. In a similar way, we find that for time $t<2d$, the reflected wave at $z=0^-$ is given by the term \eqref{fresnelr} substituted for $S$ in \eqref{invlapl}. Now we can clearly take the limit $d\to\infty$. This will not alter \eqref{fresnel}, but will ensure the validity of the associated time-domain solutions for all times. The solution has now been found; however, the evaluation of the fields \eqref{fresnel} far away in the upper half-plane is impractical. Fortunately, by Cauchy's integral theorem we can move the line $\gamma$ down towards the first branch point of $n$ or zero of $\eta+1$. If there are no such points in the closed upper half-plane, we can even evaluate \eqref{fresnel} on the real axis. Note that although the original functions $R$ and $S$ from \eqref{slabrt} are not analytic in the upper half-plane, the terms \eqref{fresnel} are analytic provided $\eta+1$ has no zeros and $n$ is analytic. Also note that in all cases, the expressions \eqref{fresnel} are analytic in some region $\im\omega>\gamma$, and $S$ tends to $\exp(i\omega z/c)$ as $\im\omega\to\infty$. In other words, $\mathcal E(z,t)=0$ for $t<z/c$ [14], and relativistic causality is guaranteed.

We conclude that the Fresnel equations \eqref{fresnel} retain their form, but not always their interpretation for active media. When $\epsilon\mu$ has no odd-order zeros in the upper half-plane, the sign of the refractive index is determined such that $n\to+1$ as $\omega\to\infty$, and such that $n$ is analytic in the upper half-plane. Furthermore, if $\eta\neq -1$ everywhere in the closed upper half-plane, the Fresnel equations \eqref{fresnel} can be evaluated and interpreted in the usual way along the real frequency axis. On the other hand, if $\epsilon\mu$ has odd-order zeros, or $\eta+1$ has zeros in the upper half-plane, \eqref{fresnel} contain points of non-analyticity. The reflected and transmitted fields \eqref{fresnel} should then be evaluated along the line $\im\omega=\gamma$, where $\gamma$ is larger than the maximum $\im\omega$ of the points of non-analyticity. The sign of the refractive index is determined so that $n\to +1$ as $\omega\to\infty$, and so that $n$ is analytic for $\im\omega\geq\gamma$. Note that when $\epsilon\mu$ has odd-order zeros in the upper half-plane, $n$ is not uniquely determined along the real frequency axis, and consequently, one must be careful with its interpretation there (See Appendix A).

For non-orthogonal incidence we can use a similar argument as that given above. When the incident wave vector is ${\bf k}=(k_x,k_y,k_z)$, the TE fields of a finite slab are given by the substitutions $\eta\to\mu k_z/k_z'$ and $k\to k_z'$ in \eqref{slabrt}. Here, $k_z'^2=\epsilon\mu\omega^2/c^2-k_x^2-k_y^2$. Again, we note that the choice of sign of $k_z'$ does not matter as only even powers of $k_z'$ are present. Following the argument leading to \eqref{fresnel} for active media, we arrive at the similar Fresnel expressions
\begin{subequations}
\label{fresneloblique}
\begin{align}
&R=\frac{\mu k_z-k_z'}{\mu k_z+k_z'},\label{fresnelroblique}\\
&S=\frac{2\mu k_z}{\mu k_z+k_z'}\exp(ik_z' z), \label{fresnelsoblique}
\end{align}
\end{subequations}
valid for $\im\omega\geq\gamma$. Here, the sign of $k_z'$ is determined such that $k_z'$ is analytic for $\im\omega>\gamma$, and such that $k_z'\to+\omega/c$ as $\omega\to\infty$. The parameter $\gamma$ is chosen larger than the maximum imaginary part of branch points of $k_z'$ or zeros of $\mu k_z+k_z'$. If there are no such points in the closed upper half-plane, $\gamma$ can be chosen to be zero, i.e., the Fresnel equations can be interpreted along the real frequency axis in the usual way.

Since the refractive index is analytic in the half-plane $\im\omega>\gamma$, and since $n-1$ has the required asymptotic behavior for $\im\omega\geq\gamma$, the Kramers-Kronig relations for active media can still be written as in \eqref{KKn}; however, the integrals are taken along the line $\im\omega=\gamma$. Note that the conventional Kramers-Kronig relations for $n$, as integrated and evaluated along the real frequency axis, are not valid.

\section{Examples}
\subsection{Inverted Lorentzian media}\label{example1}
In this example, we consider an inverted Lorentzian medium with $\epsilon=1-f$, where $f$ is given by \eqref{lorentz}, and $\mu=1$. Since $f\neq 0$ everywhere, $\eta+1$ has no zeros. Conventional optical gain media, such as the Erbium-doped silica fiber, have relatively low gain per length of fiber. We will therefore first consider the well-known case where $f$ satisfies $|f|\ll 1$. Then $\epsilon$ has no zeros, and following the recipe above, we find that the Fresnel equations \eqref{fresnel} can be interpreted straightforwardly for real frequencies, and $n\approx 1-f/2$.

Lifting the assumption of low gain, we observe that the zeros of $\epsilon$ are located at $\omega_{\pm}=-i\Gamma/2\pm i\sqrt{\Gamma^2/4+(F-1)\omega_0^2}$. When $F>1$, the zero $\omega_{+}$ is located in the upper half-plane. Consequently, in this case neither $n$ nor the Fresnel equations have unique physical meaning along the real frequency axis. The fields should instead be evaluated along the line $\im\omega=\gamma$, where $\gamma$ is chosen such that $\gamma>\im\omega_{+}$.

It is useful to recall that the fields, as given for $\im\omega>\im\omega_+$, can be interpreted by choosing an exponentially increasing excitation. For example, we may choose an excitation $\mathcal E^+(0,t)=u(t)\exp(-i\omega_1 t)$, where $\im\omega_+<\im\omega_1<\gamma$. For sufficiently large $t$, the time-domain field \eqref{invlapl} gets the main contribution from the residue of $S\exp(-i\omega t)/(i\omega_1-i\omega)$ at $\omega_1$. In other words, $S$ is roughly the complex amplitude of the exponentially increasing wave. The resulting field dependence $\exp(in\omega_1 z/c-i\omega_1 t)$ means that the wave has its ``phase velocity'' in the positive $z$-direction. (For this medium $\re n>0$ in the first quadrant of complex frequency.)

The exponentially increasing excitation is somewhat artifical, and it would be interesting to evaluate the time-domain solution when $\im\omega_1<0$, i.e., when the excitation pulse envelope is exponentially decreasing. In Appendix A, the time-domain field at a fixed $z$ is evaluated, and it is argued that the pulse diverges as $t\to\infty$. Note that the field in the medium diverges as a result of the medium alone, not as a result of, for instance, amplified, multiple reflections. 

\subsection{Excitation of a ``backward'' wave}\label{example2}\label{backexcitation}
Consider the medium $\epsilon=(1+f)^2$ and $\mu=1$, where $f$ is still given by \eqref{lorentz}. The associated susceptibilities satisfy \eqref{KKchi} and \eqref{sym}, so at least in principle, this medium is realizable. The product $\epsilon\mu$ is zero-free in the upper half-plane, and we can determine the refractive index such that it is analytic in the upper half-plane (and such that $n\to+1$ as $\omega\to\infty$). This gives $n=1+f$ and $\eta=1/(1+f)$. Note that $\eta\neq-1$ for $\im\omega\geq 0$. It follows that the Fresnel equations \eqref{fresnel} can be interpreted straightforwardly along the real frequency axis.

When $F$ is sufficiently large, there exists a real frequency $\omega_1=\omega_0\sqrt{1+F/2}$ with $\re n(\omega_1)\approx -1$ while $\im n(\omega_1)\approx \frac{4\Gamma}{F\omega_0}\sqrt{1+F/2}$ is small. This seems to be a potential route to negative refractive index at optical frequencies \cite{chen2005}. Since $\re\epsilon(\omega_1)\approx 1$ and $\mu=1$, not only the phase velocity but also the Poynting vector point in the negative $z$-direction at this frequency. As $\im n(\omega_1)>0$ and $\im\epsilon(\omega_1)<0$, the field can be interpreted as a growing wave propagating in negative $z$-direction. Nevertheless, the excitation of this ``backward'' wave does neither rely on finiteness of the thickness of the active medium nor excitation from $z=+\infty$. Note that any causal excitation involves an infinite band of frequencies, and outside the limited negative-index band, the phase velocity and Poynting vector point in the forward direction.

\begin{figure}[t!]
\includegraphics[height=5.5cm,width=7.5cm]{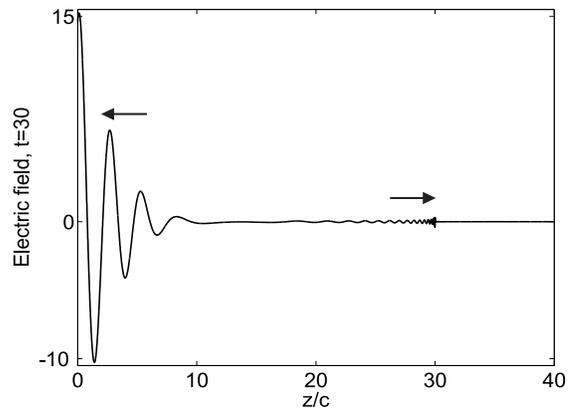}
\caption{The electric field $\mathcal E(z,t)$ of the example in Subsection \ref{backexcitation}. The frequency $\omega_0=1$ (normalized), and $t=30$. The parameters used are $\Gamma/\omega_0=0.1$ and $F=10$. The field was calculated by a straightforward numerical evaluation of \eqref{invlaplex2} using $\gamma=0.01\omega_0$. The field plot for small $z$ shows the ``backward wave'', with phase velocity and Poynting's vector in the $-z$-direction. The forerunner, located close to $z=30c$, moves at a speed $c$ in the $+z$-direction.}
\label{fig:backwave}
\end{figure}

For this medium, the transmitted time-domain field at $z=0^+$ can be calculated analytically. In this case we choose a real excitation of the form $\mathcal E^+(0,t)=u(t)\cos(\omega_1 t)$, where $\omega_1=\omega_0\sqrt{1+F/2}$. With the help of the corresponding Laplace transform and the Fresnel transmission coefficient $2/(n+1)$, we find after some simple algebra:
\begin{equation}
\label{invlaplex2}
\mathcal E(0,t)=\frac{i}{2\pi}\int_{i\gamma-\infty}^{i\gamma+\infty} \frac{(\omega^2+i\omega\Gamma-\omega_0^2)\omega\exp(-i\omega t)}{(\omega-\omega_2)(\omega-\omega_{-2})(\omega^2-\omega_1^2)}\diff\omega.
\end{equation}
Here $\gamma=0^+$, and $\omega_{\pm 2}=\pm\omega_1-i\Gamma/2$ to first order in $\Gamma/\omega_1$. The inverse transform \eqref{invlaplex2} can be found exactly by calculating the residues in the closed, lower half-plane. Assuming $\Gamma/\omega_1\ll 1$, this procedure yields
\begin{equation}
\label{invlaplex2f}
\mathcal E(0,t)=
\begin{cases}
\cos(\omega_1 t), & \text{for $0<\Gamma t/2\ll 1$}, \\
-\frac{2}{\im n(\omega_1)}\sin(\omega_1 t), & \text{for $\Gamma t/2\gg 1$}.
\end{cases}
\end{equation}
(The steady-state solution could certainly be found directly from \eqref{fresnels}.) We observe that the transmitted field is equal to the excitation for small $t$. Thus, initially the reflection from the boundary vanishes. On the other hand, when $t$ is large, the field has grown considerably. This fact is interpreted by noting that the ``backward wave'', with phase velocity and Poynting's vector in the negative $z$-direction, has been built up. This wave draws energy from the active medium, yielding a large field at the boundary. Note that also the ``reflected wave'' on the left-hand side of the boundary now is large. Nevertheless, the fields are produced in a stable manner. A plot of the backward wave and the associated, forward propagating forerunner is given in Fig. \ref{fig:backwave}.
 
At oblique incidence the backward wave is refracted negatively. Although this may seem obvious intuitively, the fact can be proved by calculating $k_z'$ and identifying its sign along the real frequency axis using the procedure outlined above. Despite negative refraction, the medium has rather different properties from the left-handed negative-index medium considered by Veselago and Pendry \cite{veselago,pendry2000}. Considering a finite slab of sufficiently small thickness such that $T$ has no poles in the upper half-plane, we find $T\approx\exp(+i\omega_1 d/c)$ at the frequency $\omega_1$. Also, evanescent-field amplification, as in the Veselago-Pendry lens, will not arise in this medium. For TE polarization this is seen by the substitutions $\eta\to\mu k_z/k_z'$ and $k\to k_z'$ in \eqref{slabt}. As before, $k_z'^2=\epsilon\mu\omega^2/c^2-k_x^2-k_y^2$. For an incident evanescent field with sufficiently large $k_y$, $k_z'$ is close to an imaginary number. Substitution into \eqref{slabt} gives exponential decay. Note that the choice of sign of $k_z'$ does not matter.

An interesting point arises by comparing the results in this example with those of the previous example. Assuming low gain in the previous example, $\epsilon\approx 1-i\alpha$ for some real frequency $\omega_1$. Here, $\alpha$ is a small positive number. In the present example, at the frequency $\omega_1$ we can also write $\epsilon\approx 1-i\alpha$ for a small positive $\alpha$. Indeed, it is possible to tune the parameters in the two examples such that the two $\epsilon$'s (and $\mu$'s) are identical at this frequency. Nevertheless, the Fresnel equations \eqref{fresnel} give completely different answers; in the previous example the ``forward'' wave, with phase velocity and energy growth in the $+z$-direction, is excited, whereas in the present example a ``backward'' wave is excited. The dilemma is resolved by noting that although the $\epsilon$'s are identical for a single frequency, they are not identical globally. As a result, the causal excitation at $z=0^-$, which necessarily contains an infinite band of frequencies, will have different consequences in the two situations.

\subsection{Instability at the boundary}\label{example3}
In the previous example, although $n\approx -1$ at the real frequency $\omega_1$, $n=-1$ is exactly fulfilled only in the lower half-plane (at the complex frequencies $\omega_{\pm 2}$). In principle, one can construct a causal, active medium for which $n=-1$ at a real frequency or even in the upper half-plane. Indeed, let $\epsilon=(1+g)^2$ and $\mu=1$, where
\begin{equation}
\label{gfunc}
g(\omega)=\frac{i G\omega}{\omega_{0}^2-\omega^2-i\omega\Gamma}.
\end{equation}
We assume $\Gamma/\omega_0\ll 1$ and $G\sim\Gamma$. The associated susceptibility $\chi=\epsilon-1$ is clearly analytic and uniformly square integrable in the closed upper half-plane; thus it satisfies the Kramers-Kronig relations \footnote{One may argue that the asymptotic form of $\epsilon-1$ for large $\omega$ is different from the usually assumed dependence $1/\omega^2$. Without significantly altering $g$ in the region $|\omega-\omega_0|\lesssim\Gamma$ this can be fixed for instance by letting $g\to (g+\tilde F)\tilde\omega_0^2/(\tilde\omega_{0}^2-\omega^2-i\omega\tilde\Gamma)$ where $|\tilde F|\ll 1$, $\tilde\omega_0\gg\omega_0$, and $0<\tilde\Gamma\ll\tilde\omega_0-\omega_0$.}. By determining the refractive index such that it is analytic in the upper half-plane, we obtain $n=1+g$. Thus we can find the frequencies $\omega_{a\pm}$ where $n=a$ is fulfilled: 
\begin{equation}
\label{freqnm1}
2\omega_{a\pm}=\pm 2\omega_0-i\Gamma-iG/(a-1).
\end{equation}
Here we have assumed that $|a-1|\gtrsim 1$ and omitted second or higher order terms in $\Gamma/\omega_0$. Using \eqref{freqnm1} we find that if $G\geq 2\Gamma$, the solutions to the equation $n=-1$ are located in the closed, upper half-plane (equality implies locations on the real axis).

Substituting \eqref{fresnels} into \eqref{invlapl}, it is realized that the locations of the zeros of $n+1$ are crucial for the electromagnetic stability. Clearly if $1/(n+1)$ has poles in the upper half-plane, the associated residues when calculating \eqref{invlapl} grow exponentially with $t$. In other words, the system consisting of the two media (vacuum and the active medium) is electromagnetically unstable. For a given active medium, this instability may be eliminated if the vacuum on the left-hand side of the boundary is replaced by another passive medium: Let the new medium have $\mu=1$ and (approximately) constant refractive index $n_{\text{p}}$ for $|\omega-\omega_0|\lesssim\Gamma$. The instability would not be present provided $n\neq -n_{\text{p}}$ in the upper half-plane. According to \eqref{freqnm1}, if $2\Gamma\leq G\leq(n_{\text{p}}+1)\Gamma$, there is no instability, in contrast to what we found above for vacuum.

\section{Conclusions}
General forms of the Fresnel equations, valid for active media, have been proved. When the refractive index contains branch points in the upper half-plane, the index is not determined uniquely along the real frequency axis, but is dependent on the choice of branch cuts. Thus the Fresnel equations and the refractive index cannot be interpreted along the real frequency axis unless their behavior along the branch cut are taken into account as well. The expressions should rather be evaluated or interpreted in some upper half-plane $\im\omega\geq\gamma$, where the line $\im\omega=\gamma$ is located above the non-analytic points. The presence of branch cuts in the upper half-plane leads generally to instabilities in the sense that any vanishing small pulse excites an infinite field (or saturation) in the medium.

The sign of the refractive index should be defined using the asymptotic form of $n$ ($n\to+1$ as $\omega\to\infty$), and analyticity for $\im\omega>\gamma$. It would have been useful to have an explicit relation for the refractive index, similar to \eqref{nfromepsilonmu}. However, for active media such a relation is not valid as the complex arguments of $\epsilon$ and $\mu$ may take any values. Nevertheless, if $\gamma$ is above all non-analytic points of $n$, $n$ is analytic for $\im\omega=\gamma$. Thus the sign can be identified up to a global sign by unwrapping the phase $\arg\epsilon+\arg\mu$ before using \eqref{nfromepsilonmu}. The global sign is determined easily by requiring $n\to +1$ as $\omega\to\infty$. 

For most media of interest, there are no branch points of $n$ for $\im\omega\geq 0$, and the sign of $n$ is to be determined along the real frequency axis. In principle, $n$ is not necessarily analytic for real frequencies (since $\epsilon$ and $\mu$ are not necessarily analytic there). With the extra assumption that $\epsilon\mu$ is continuous for real frequencies, $\epsilon\mu$ is continuous for $\im\omega\geq 0$. Then the sign of $n$ may still be found by unwrapping the phase $\arg\epsilon+\arg\mu$ in \eqref{nfromepsilonmu}, and requiring $n\to +1$ as $\omega\to\infty$.

Using the general framework, we have seen that in certain active media, only the ``backward'' propagating wave is excited. For this wave both phase velocity and Poynting's vector point towards the excitation source. These media are conveniently described using a negative refractive index. Nevertheless, evanescent-field amplification, similar to that in the Veselago-Pendry lens, does not exist.

\appendix
\section{Transmitted field when $n$ has branch points}
I will here discuss the time-domain field \eqref{invlapl} associated with the solution \eqref{fresnels}, in order to provide an interpretation of media where $n$ has branch points in the upper half-plane. For simplicity the medium is chosen to be of the type discussed in Section \ref{example1}: The electromagnetic parameters are $\epsilon=1-f$ and $\mu=1$, where $f$ is given by \eqref{lorentz}, and $F>1$. This leads to a single branch point of $n$ at $\omega=\omega_+$, where $\re\omega_+=0$ and $\im\omega_+>0$. 

\begin{figure}[t!]
\includegraphics[height=5.5cm,width=7.5cm]{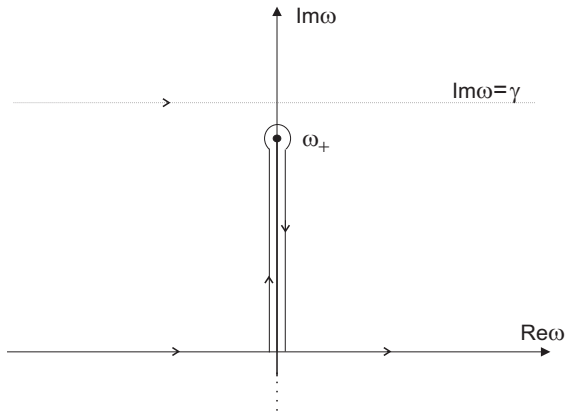}
\caption{Branch cut and integration paths. The Bromwich path $\im\omega=\gamma$ can be deformed into a path consisting of two contributions: Integration along the real frequency axis and integration around the branch cut.}
\label{fig:integrationpath}
\end{figure}

The branch cut is chosen along the imaginary frequency axis, from the branch point at $\omega_+$ towards $-i\infty$, see Fig. \ref{fig:integrationpath}. With this choice, the time-domain field can clearly be calculated by locating the Bromwich path in \eqref{invlapl} above $\omega_+$  ($\gamma>\im\omega_+$):
\begin{equation}
\label{invlaplex1}
\mathcal E(z,t)=\frac{1}{\pi}\int_{i\gamma-\infty}^{i\gamma+\infty} \frac{\exp(in\omega z/c)\exp(-i\omega t)}{(n+1)(i\omega_1-i\omega)}\diff\omega.
\end{equation}
It would however be interesting to locate the integration path along the real axis, to investigate the interpretation of the refractive index for real frequencies. Then, to obtain a correct result, we must add an integral around the branch cut. That is,
\begin{equation}
\label{invlaplbranchpoint}
\mathcal E(z,t)=\mathcal E_{\mathcal{F}}(z,t)+ \Delta\mathcal E(z,t),
\end{equation}
where $\mathcal E_{\mathcal{F}}(z,t)$ is the contribution from the integration along the real axis (inverse Fourier transform), and $\Delta\mathcal E(z,t)$ results from the integration around the branch cut (see Fig \ref{fig:integrationpath}). Since $n$ is bounded for $\im\omega\geq 0$, the contribution from the integration along the circle centered about $\omega_+$ can be neglected when the radius approaches zero. Thus we obtain
\begin{eqnarray}
\Delta\mathcal E(z,t)&&=\frac{1}{\pi}\int_{0}^{\im\omega_+}\frac{\exp(\omega_i t)}{\omega-\omega_1}\\
&&\cdot\left[\frac{\exp(n\omega_iz/c)}{n-1}+\frac{\exp(-n\omega_iz/c)}{n+1}\right]\diff\omega_i\nonumber, 
\end{eqnarray}
where $n=n(i\omega_i)$ is evaluated along the right-hand side of the branch cut. For example, considering small $z$ such that $|n\omega_i z/c|\ll 1$ along the branch cut, it is clearly seen that $\Delta\mathcal E(z,t)\to\infty$ as $t\to\infty$. Since the integrand in \eqref{invlaplex1} is square integrable along the real axis, so is $\mathcal E_{\mathcal F}(z,t)$. Consequently, the electric field $\mathcal E(z,t)$ diverges as $t\to\infty$. 

The branch cut in Fig. \ref{fig:integrationpath} could of course be chosen in a different manner, as long as it is located under the line $\im\omega=\gamma$. For each choice of branch cut, the sum $\mathcal E_{\mathcal{F}}(z,t)+ \Delta\mathcal E(z,t)$ will be the same; however the two separate terms will generally be different if the branch cut crosses the real axis at a different point. This shows that the inverse Fourier transform part $\mathcal E_{\mathcal{F}}(z,t)$ has no physical meaning unless it is seen together with $\Delta\mathcal E(z,t)$. In other words, the refractive index has no unique, physical interpretation along the real frequency axis unless its properties along the branch cut are taken into account.

\bibliography{causbib}

\end{document}